\documentclass[%
 aip,
 amsmath,amssymb,
 reprint,%
]{revtex4-1}

\usepackage{graphicx}
\usepackage{dcolumn}
\usepackage{bm}
\usepackage{ulem}
\usepackage[utf8]{inputenc}
\usepackage[T1]{fontenc}
\usepackage{mathptmx}
\usepackage{subfigure}
\usepackage{etoolbox}
\usepackage{dsfont}
\usepackage{amsthm}
\usepackage{cleveref}
\usepackage[ruled,linesnumbered]{algorithm2e}

\usepackage{mathtools}
\usepackage{tikz}
\usepackage{float}
\usepackage{braket}

\newcommand{\REV}[1]{{#1}}
\newcommand{\REVV}[1]{{#1}}

\RestyleAlgo{ruled}

\makeatletter
\def\@email#1#2{%
 \endgroup
 \patchcmd{\titleblock@produce}
  {\frontmatter@RRAPformat}
  {\frontmatter@RRAPformat{\produce@RRAP{*#1\href{mailto:#2}{#2}}}\frontmatter@RRAPformat}
  {}{}
}%
\makeatother
\begin{document}

\preprint{AIP/123-QED}
\title{Rethinking Noise in Quantum Machine Learning: When Noise Improves Learning}

\author{Linghua Zhu}
\altaffiliation{Authors contributed equally}
\affiliation{Department of Chemistry, University of Washington, Seattle, WA 98195, U.S.A.}

\author{Ziyu Zhang}
\altaffiliation{Authors contributed equally}
\affiliation{Department of Chemistry, University of Washington, Seattle, WA 98195, U.S.A.}

\author{Yulong Dong}
\altaffiliation{Corresponding author}
\affiliation{Department of Electrical Engineering and Computer Science, University of Michigan, Ann Arbor, MI 48109, U.S.A.}
\email{dongyl@umich.edu}

\author{Xiaosong Li}
\altaffiliation{Corresponding author}
\affiliation{Department of Chemistry, University of Washington, Seattle, WA 98195, U.S.A.}
\email{xsli@uw.edu}

\date{\today}

\begin{abstract}
Quantum noise is conventionally viewed as a fundamental obstacle in near-term quantum computing, motivating extensive error correction and mitigation strategies. \REV{We present numerical evidence within an effective noise modeling framework that challenges this consensus. Through experiments on quantum graph neural networks for molecular property prediction, we observe heterogeneous, initialization-dependent responses within this effective noise modeling framework.} Among randomly initialized models with identical architecture, approximately one-third show performance improvement under moderate noise, while a smaller fraction deteriorate and the remainder are marginally affected. We identify a strong negative correlation ($r = -0.62$) between baseline model performance and noise benefit, suggesting \REV{a regularization-like effect for under-optimized models while disrupting well-converged ones}. The observed optimal noise level falls below theoretical predictions, indicating error cancellation in structured quantum circuits. These findings \REV{suggest that, within the adopted effective noise model, noise effects depend critically on initialization quality and need not be uniformly detrimental}, motivating structure- and noise-aware optimization strategies.

\end{abstract}

\maketitle

\section{Introduction}

Noisy intermediate-scale quantum (NISQ) devices operate with gate error rates on the order of $10^{-3}$-$10^{-2}$, which fundamentally constrains their practical utility to variational algorithms designed to tolerate imperfect operations.
Recent studies have begun to reveal that quantum noise, traditionally viewed as a limiting factor, can in some cases play a constructive role across a range of computational paradigms, leading to unexpected performance benefits.

Du et al.~\cite{Liu21_023153} showed that depolarizing noise can provide intrinsic robustness against adversarial attacks through a mechanism analogous to differential privacy. In the context of quantum reservoir computing, amplitude-damping noise has been found to enhance performance when the gate depth and error rates lie within specific regimes \cite{ Borondo23_8790}. Stochastic noise has also been demonstrated to assist variational algorithms in escaping saddle points and shallow local minima \cite{ Eisert25_052441}. More recently, noise itself has been proposed as a computational resource for simulating open quantum systems, leveraging intrinsic decoherence rather than treating it solely as an error source \cite{ Zhang25_251021075}.

The potential for noise to act as an implicit regularizer has attracted growing attention. Somogyi et al. demonstrated that treating noise strength as a tunable hyperparameter can enhance quantum neural network performance by up to 8\% on clinical prediction tasks \cite{ Melnikov25_e00603}. Scala et al. further proposed that optimal noise levels induce parameter equalization, whereby less influential parameters gain relevance while previously dominant ones are suppressed \cite{ Lucchi26_251109428}. While these studies highlight the constructive role noise can play, the extent to which such benefits depend on the underlying noise model and problem structure in practical quantum machine learning applications remains largely unexplored.

While these studies demonstrate that quantum noise can improve performance under certain conditions, they leave fundamental questions unanswered. It remains unclear why some models benefit from noise while others degrade, or whether such responses can be predicted from readily observable model characteristics. Most existing work evaluates noise effects by averaging across multiple training runs, potentially masking important variations in how individual models respond. A systematic understanding of what determines whether noise helps or harms a given quantum machine learning model is still lacking. Without such understanding, practitioners cannot make informed decisions about when to invest resources in aggressive error mitigation versus when to tolerate or even leverage existing noise levels.

In this work, we present a numerical study of the response of quantum graph neural networks (QGNNs) to quantum noise in molecular property prediction, with particular emphasis on model initialization.
We train 55 independently initialized QGNN models on a subset of QM9 dataset and analyze each model's response to four noise levels spanning from noiseless conditions to typical NISQ hardware error rates. By examining individual models rather than population averages, we reveal that noise response is not uniform but varies systematically with baseline model performance. Models that converge to lower quality solutions tend to benefit from noise, while better optimized models experience degradation. This systematic relationship enables prediction of noise response from baseline characteristics and points toward structure- and noise-aware optimization strategies for quantum machine learning in the NISQ era.

\section{METHODS}
\label{sec:method}
\subsection{Dataset and Task}
In our study, we randomly sample 2,000 molecules from the QM9 dataset, which contains quantum chemical properties for approximately 134,000 organic molecules in total\cite{ VonLilienfeld14_140022}. Each molecule contains up to nine heavy atoms drawn from carbon, nitrogen, oxygen, and fluorine. Molecular structures are represented as undirected graphs, with nodes corresponding to atoms and edges representing chemical bonds of different types (single, aromatic, double, or triple). We focus on predicting the energy gap between the highest occupied molecular orbital (HOMO) and the lowest unoccupied molecular orbital (LUMO). The HOMO-LUMO gap is a fundamental quantum mechanical property that governs molecular reactivity, optical behavior, and electronic response, and is therefore a widely used and physically meaningful benchmark for evaluating quantum machine learning models.

Following standard protocol, we randomly split the dataset into training (80\%), validation (10\%), and test (10\%) sets, maintaining consistent splits across all experiments.

We evaluate model performance using the coefficient of determination ($R^2$ score). To quantify noise effects, we calculate the relative performance change as 
\begin{equation}
    \Delta R^2 = \frac{R^2_{\mathrm{noisy}} - R^2_{\mathrm{noiseless}}}{R^2_{\mathrm{noiseless}}} \times 100\%.\label{eq:change}
\end{equation}
Because a 2\% change in $R^2$ is comparable to the typical variance observed across training runs, we classify model responses to noise as positive ($\Delta R^2 > 2\%$), negative ($\Delta R^2 < -2\%$), or marginal (changes within $\pm 2\%$).

\subsection{Quantum Graph Neural Network Architecture}

The QGNN implementation builds upon the equivariantly diagonalizable unitary (EDU) quantum graph circuits framework, following the architecture developed by Ryu et al. \cite{ Rhee23_4300} for molecular property prediction in materials science. 
The EDU approach, originally formalized by Mernyei et al. \cite{ Ceylan22_15401}, guarantees permutation equivariance with respect to node ordering, a critical requirement ensuring that predictions remain invariant under different atom labeling schemes. Unlike generic parameterized quantum circuits that may waste computational resources learning trivial symmetries, the EDU framework incorporates these constraints directly into the circuit architecture, reducing the effective parameter space while maintaining expressiveness \cite{Eisert23_010328}.

The circuit employs 12 qubits, including 11 encoding qubits and 1 master qubit,\REV{ which provides an optional graph-level readout channel that is not activated in the present study,} applicable to all QM9 molecules. In the experiments reported in this work, the atom-encoding qubits form a fixed atom-register space, which encodes atomic information and processes local interactions. Hydrogen atoms are not explicitly represented as graph nodes. Therefore, each molecule is represented as a heavy-atom graph constructed from the RDKit molecule structure. For a molecule containing $N$ heavy atoms, with $N \leq 9$ in the QM9 subset used here, the first $N$ qubits are assigned to the heavy atoms according to the RDKit atom ordering. The remaining $11 - N$ qubits did not correspond to atoms and served as inactive padding qubits, which remained initialized in $\lvert 0 \rangle$ and did not receive atom-encoding gates, allowing molecules of different sizes to share the same fixed-size quantum register and fixed-dimensional quantum feature vector.

\REV{ Since all molecules considered here contain at most 9 heavy atoms, the master qubit remains in its initial state during the present experiments, and its Pauli-$Z$ expectation value, together with those of the padding qubits, is constant and contributes only a constant feature to the classical network. The master qubit is reserved for a master-node extension of the EDU-QGC framework, in which a dedicated node connected to all atom qubits performs graph-level readout and this extension is not enabled here, the graph-level aggregation is performed by the classical readout network. A 9-qubit register would therefore suffice for this task \REVV{ since the idle qubits remain in their initial state and their Pauli-$Z$ expectation values are constant}. The larger register is an implementation choice rather than a requirement of the EDU-QGC construction, with additional qubits available for larger molecules and for the optional master-node readout.}

Each active atom qubit was initialized from $\lvert 0 \rangle$ and encoded using element-dependent rotation parameters. Specifically, the element type of atom $i$ was denoted as $s_i$, where $s_i$ corresponds to the atomic symbol. Each element type corresponds to a different set of rotation parameters $[\alpha_{s_i}, \beta_{s_i}]$, which control the base angle for RY and RZ rotation, respectively. To include local chemical environment information, the RY angle is scaled by the atomic degree through the factor $\alpha_{s_i}(1 + d_i/4)$, where $d_i$ is the degree of atom $i$. The RZ angle is scaled by aromaticity through the factor $\beta_{s_i}(1+ A_i)$, where $A_i$ is set to 0.5 for aromatic atoms and 0 otherwise. With this encoding method, the active qubits combine not only the atom identity but also the local connectivity and aromaticity information.
\cite{Coles21_6961}



\begin{figure}[htbp]
\centering
\includegraphics[width=\linewidth]{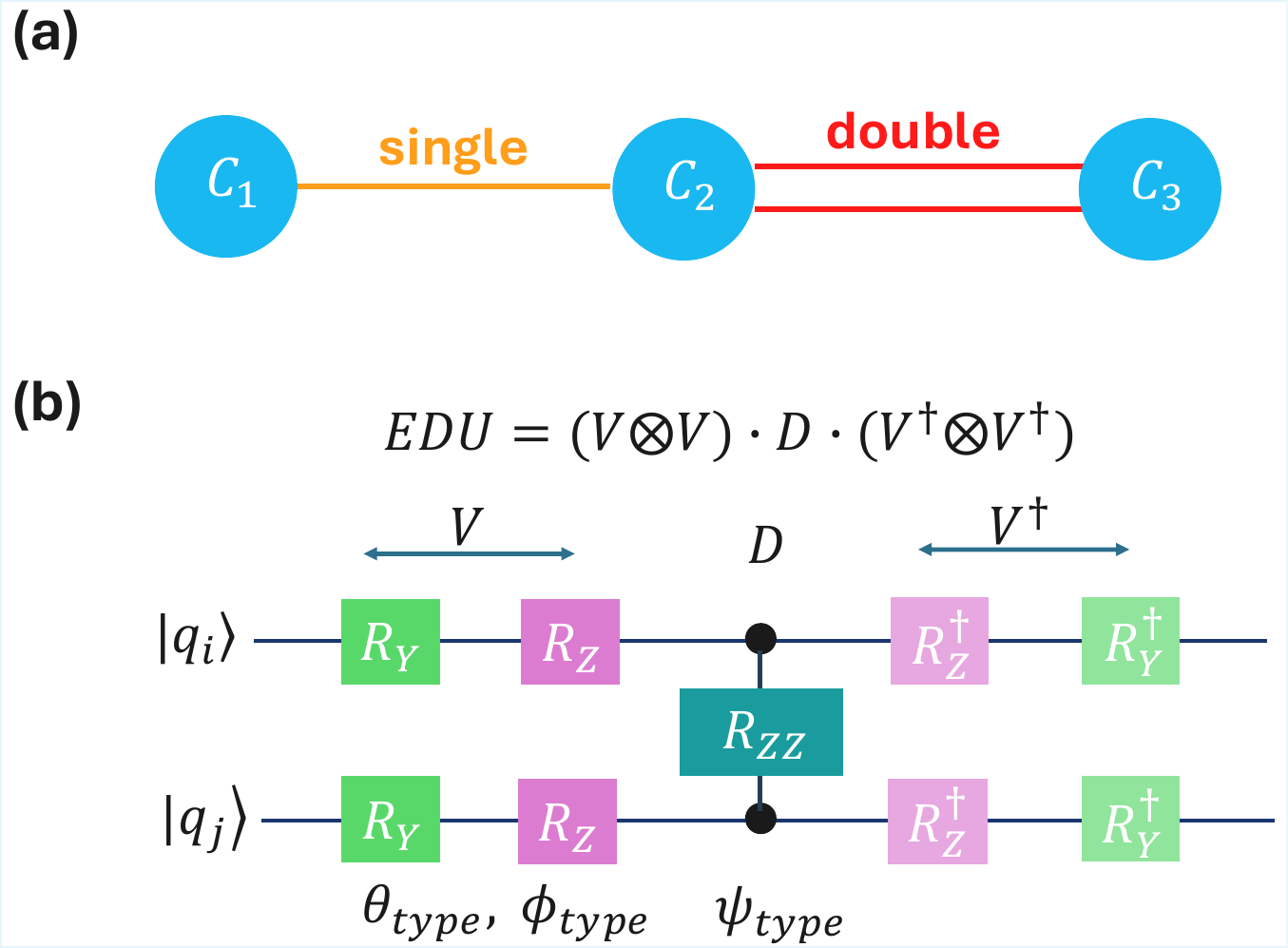}
\caption{EDU bond encoding mechanism. (a) Molecular graph representation showing nodes (atoms) and edges with different bond types (single and double bonds). (b) EDU circuit structure. The quantum circuit implements the decomposition $\mathrm{EDU} = (V \otimes V) \cdot D \cdot (V^\dagger \otimes V^\dagger)$, where node-local unitaries $V$ use bond-type-specific parameters ($\theta_{\mathrm{type}}, \phi_{\mathrm{type}}$) and the diagonal unitary $D$ uses layer-shared parameters ($\psi_{\mathrm{layer}}$). }
\label{fig:edu_encoding}
\end{figure}

The EDU operations process molecular bonds according to their chemical type, with parameters learned separately for single, double, triple, and aromatic bonds. Each molecular bond triggers an EDU operation implementing quantum message passing between connected atoms. 
The EDU performs three transformations: encoding bond features via parameterized rotations $R_Y(\theta_{ij})$ and $R_Z(\phi_{ij})$, creating entanglement between connected atoms through $R_{ZZ}(\psi_{ij})$ gates, and decoding to preserve unitarity \cite{ Ceylan22_15401}, as illustrated in \cref{fig:edu_encoding}. This structure ensures that entanglement reflects the chemical graph topology while maintaining differentiability for gradient-based optimization~\cite{Yang25_026110}.

Circuit measurements employ Pauli-Z expectation values on all 12 qubits, producing a quantum feature vector that is processed through a classical neural network to predict the HOMO-LUMO gap. 
For a molecular graph $G$ with trainable circuit parameters $\theta$, the quantum circuit generates a feature vector through measurements:
\begin{equation}
\mathbf{z}(G, \theta) = \left[ \langle \psi_G(\theta) | Z_i | \psi_G(\theta) \rangle \right]_{i=1}^{12} \in \mathbb{R}^{12} 
\end{equation}
\REV{where $|\psi_G(\theta)\rangle = U_{\rm EDU}(G,\theta)\, U_{\rm node}(\theta)\, U_{\rm enc}(G,\theta)\,|0\rangle^{\otimes 12}$ with $U_{\rm enc}(G,\theta)$ denoting the element-dependent encoding rotations described above, $U_{\rm node}(\theta)$ a layer of general single-qubit rotations applied to each atom-encoding qubit, $U_{\rm EDU}(G,\theta)$ the bond-triggered EDU operations that encode the molecular graph structure, and $|0\rangle^{\otimes 12}$ the initial state.} This 12-dimensional quantum feature vector is then processed by a classical feedforward neural network to produce the final prediction:
\begin{equation}
f(G, \theta) = \text{NN}_{\text{classical}}(\mathbf{z}(G, \theta)).
\end{equation}
where $\text{NN}_{\text{classical}}$ is a 4 layer feedforward network with hidden dimensions \REVV{64, 32, and 16}, using ReLU activations and dropout regularization. This hybrid quantum-classical approach follows the variational algorithm paradigm, where quantum circuits generate features while classical networks perform the final regression \cite{Coles21_625}.

Each experimental run uses a different random initialization for all trainable parameters. Rather than averaging across runs, we analyze each initialization separately to examine how noise response varies with model characteristics. 

\subsection{Noise Model}
\label{sec:noise_model}

Quantum processors exhibit multiple decoherence mechanisms arising from coupling to environmental degrees of freedom, including energy relaxation (T1) and dephasing (T2) \cite{Oliver19_021318}. While full characterization of these processes is essential for gate-level error correction, algorithm-level studies can employ simplified effective models that capture statistical effects rather than microscopic details. Randomized benchmarking experiments demonstrate that composite noise in real devices, when averaged over many gate sequences, often exhibits effective depolarizing-like behavior \cite{Steffen12_080505, Wallman20_1184}, supporting the use of phenomenological models for studying algorithm performance.

We adopt such a phenomenological model that captures the cumulative effect of gate errors across all gates within the single-layer circuit:
\begin{equation}
f_{\text{noisy}}(G, \theta, \varepsilon) = (1 - p_{\text{error}}(\varepsilon)) \cdot f_{\text{noiseless}}(G, \theta) + \xi
\end{equation}
where the cumulative error probability follows:
\begin{equation}
p_{\text{error}}(\varepsilon) = 1 - (1 - \varepsilon)^{N_g \cdot L}\label{eq:error}
\end{equation}
Here $\varepsilon$ is the per-gate error rate, $N_g$ is an effective gate count parameter, and $L$ denotes the circuit depth (number of layers). This formulation assumes independent error accumulation across gates, a standard approximation for estimating circuit-level fidelity \cite{Chuang02_Book}. \REV{ The stochastic term $\xi \sim \mathcal{N}(0,\sigma_{\rm noise}^2)$, with $\sigma_{\rm noise} = 0.2\,p_{\rm error}$, represents residual fluctuations not captured by the systematic attenuation. The coefficient was set so that these fluctuations scale with the accumulated error probability while staying small compared with it.}

We examine four noise levels: $\varepsilon \in \{0.000, 0.005, 0.010, 0.015\}$, corresponding to per-gate fidelities of 100\%, 99.5\%, 99\%, and 98.5\% respectively. The lowest nonzero value ($\varepsilon = 0.005$, corresponding to 99.5\% fidelity) is comparable to current hardware capabilities. State-of-the-art two-qubit gate fidelities have reached 99.5\% on neutral-atom arrays \cite{Lukin23_268}, 99.9\% on trapped-ion \cite{Lucas16_060504} and superconducting \cite{Vepsalainen25_250816437} platforms, with recent advances achieving 99.99\% on trapped-ion systems \cite{Sutherland25_251017286}.

\section{Results and Discussion}

\begin{figure}[htbp]
\centering
\includegraphics[width=\linewidth]{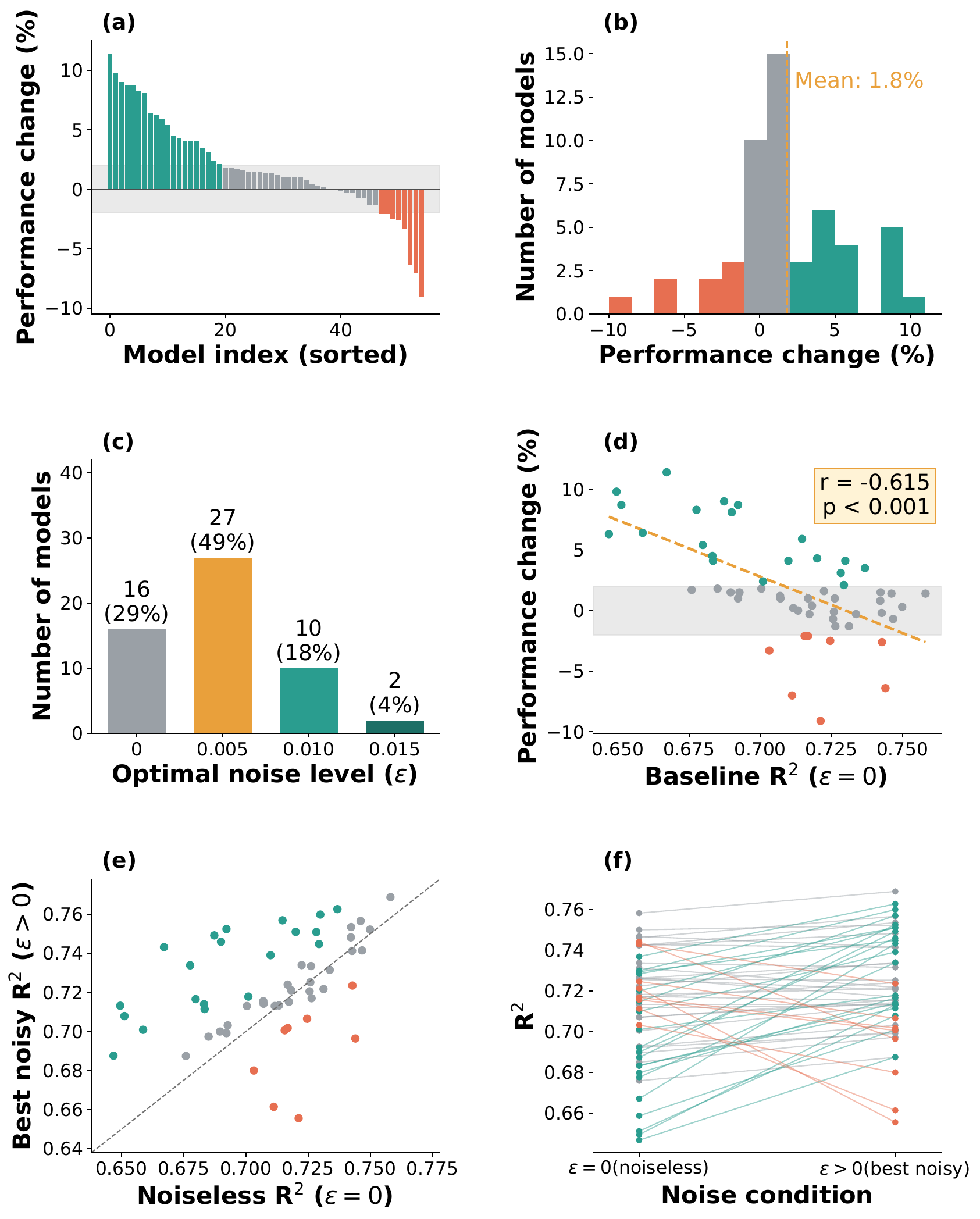}
\caption{Heterogeneous response to noise across 55 random initializations. (a) Waterfall plot of individual model responses sorted by performance change. (b) Distribution of performance changes, revealing 36.4\% beneficial ($>2\%$), 14.5\% detrimental ($<-2\%$), and 49.1\% marginal responses. (c) Optimal noise level distribution, with 49\% of models performing best at $\varepsilon = 0.005$. (d) Negative correlation ($r = -0.615$, $p < 0.001$) between baseline performance and noise benefit. \REV{(e) Best noisy test $R^2$ ($\varepsilon > 0$) against noiseless test $R^2$ ($\varepsilon = 0$); the dashed line marks $y = x$, so that points above it improve under noise and points below degrade. (f) The same values shown as paired points for each model. In (a), (b), (d), (e) and (f), teal, orange and grey denote the beneficial, detrimental and marginal groups, respectively.} }
\label{fig:heterogeneous}
\end{figure}

We examined the effect of quantum noise on single-layer QGNN performance using a systematic study of 55 independently initialized models trained on the QM9 HOMO-LUMO gap prediction task. Each model was initialized with a distinct random seed, and the same set of 55 initializations was used consistently across all noise profiles, enabling a controlled, like-for-like comparison of noise effects.

Contrary to the conventional expectation that quantum noise uniformly degrades performance, we observe a strongly heterogeneous response across the 55 models. For each model, we identify its optimal noisy performance, defined as the maximum $R^2_{\text{noisy}}$ across the four noise profiles, and report the corresponding $\Delta R^2$ (\cref{eq:change}) in \cref{fig:heterogeneous}(a). The resulting heterogeneous response spans a wide range, from a 9.1\% decrease to an 11.4\% increase when quantum noise is introduced, relative to the noiseless baseline, yielding a total spread of 20.5 percentage points. This spread exceeds five standard deviations of the typical training variance and cannot be explained by stochastic fluctuations alone (permutation test, $p < 0.001$, where $p$ denotes the probability of observing this result under the null hypothesis).

As shown in \cref{fig:heterogeneous}(c), 49\% of the models achieve their optimal performance at a noise strength of $\varepsilon = 0.005$. Notably, a substantial subset of models continues to benefit from stronger noise: 18\% and 4\% of the models reach their respective optima at $\varepsilon = 0.010$ and $\varepsilon = 0.015$, respectively, highlighting the non-monotonic and model-dependent nature of the noise response.
In contrast, 16 models (29\%) exhibit degraded performance relative to their noiseless baselines, indicating that quantum noise remains detrimental for a significant number of tests.

If we assume that noise benefit is maximized when cumulative error probability reaches $p_{\text{error}} = 0.5$, the theoretical optimal per-gate error rate can be derived from \cref{eq:error}:
\begin{equation}
\varepsilon_{\text{optimal}}^{\text{theory}} \approx \frac{\ln 2}{N_g \cdot L}
\end{equation}
For our single-layer architecture ($L=1$) with 50 to 100 gates for the QM9 set, this yields $\varepsilon_{\text{optimal}}^{\text{theory}} \approx 0.007$ to $0.014$. However, the experimentally observed optimal of $\varepsilon = 0.005$ is considerably lower, suggesting effective error cancellation in structured quantum circuits. This discrepancy may arise from symmetry constraints inherent to the EDU architecture or correlated errors that partially cancel rather than accumulating independently. Characterizing these error cancellation mechanisms remains an important direction for future work.

Across all 55 models, the mean improvement is 1.8\% (\cref{fig:heterogeneous}(b)). Noise-beneficial cases exhibit larger average gains (5.8\% $\pm$ 2.9\%) than the average magnitude of degradation observed in noise-detrimental cases (4.2\% $\pm$ 2.1\%), producing a clear asymmetry in the response distribution and indicating that noise-induced improvements are typically more pronounced than noise-induced degradations.
Of the 55 models tested, 36.4\% demonstrated statistically significant performance improvements ($>$2\% increase in $R^2$) when quantum noise is introduced, while only 14.5\% experienced substantial degradation ($<-2\%$ decrease). The remaining 49.1\% exhibited marginal responses within $\Delta R^2=\pm$2\%. 

\Cref{fig:heterogeneous}(d) reveals a clear negative correlation between baseline model performance ($R^2$ at $\varepsilon = 0$) and the noise-induced performance change $\Delta R^2$, indicating that better-performing noiseless models tend to benefit less or even degrade when quantum noise is introduced, whereas weaker baseline models are more likely to gain from moderate noise.  Using a threshold at $R^2 = 0.71 (\varepsilon=0)$ (near the midpoint between beneficial and detrimental group baseline means), models below this value show a 61\% probability of benefiting from noise, while those above show a 22\% probability of degradation.

\REV{\Cref{fig:heterogeneous}(e) and (f) show the same results in absolute terms. The noiseless $R^2$ values range from 0.647 to 0.758 and the best noisy values from 0.656 to 0.769, so that the relative changes above correspond to shifts of up to about 0.08 in $R^2$.}

\begin{figure}[htbp]
\centering
\includegraphics[width=\linewidth]{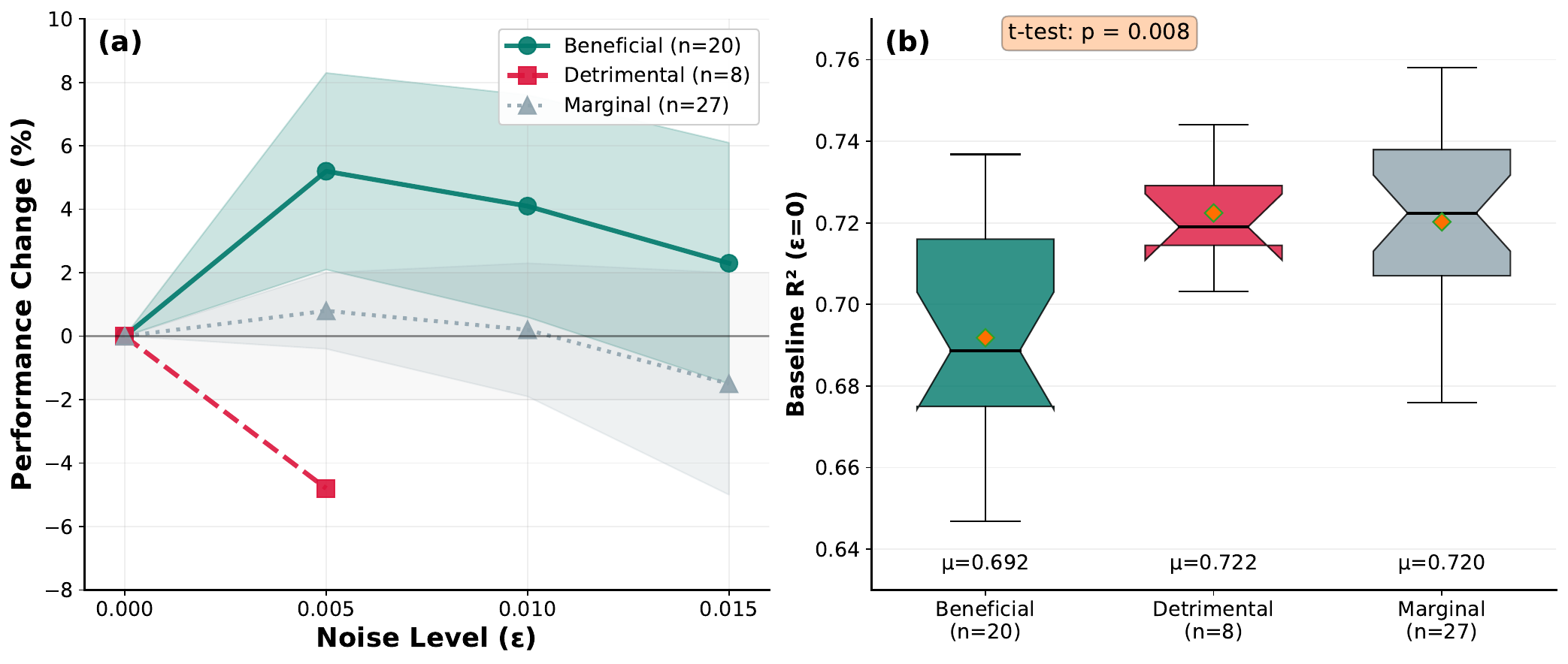}
\caption{Dose-response analysis and mechanistic basis. (a) Average performance change as a function of noise level for beneficial, detrimental, and marginal groups. (b) Baseline performance distribution across response categories reveals systematic differences.}
\label{fig:mechanism}
\end{figure}

Analysis of dose-response relationships reveals distinct patterns among the three response categories (\cref{fig:mechanism}(a)). The noise-beneficial group reaches maximum improvement at $\varepsilon = 0.005$ before declining at higher noise levels. In contrast, the noise-detrimental group shows immediate performance degradation, with 75\% triggering early stopping protocols at $\varepsilon = 0.005$. The marginal group maintains relatively stable performance across all noise levels.

Models in the beneficial category exhibit significantly lower baseline $R^2$ scores (0.692 $\pm$ 0.028) compared to the detrimental category (0.722 $\pm$ 0.014; two-sample $t$-test, $p = 0.008$; \cref{fig:mechanism}(b)). This separation in baseline performance between response categories supports the interpretation that initial optimization quality is a primary factor determining noise sensitivity. Using a threshold at $R^2 = 0.71$, models below this value show a 61\% probability of benefiting from noise, while those above show a 22\% probability of degradation. While this relationship is probabilistic rather than deterministic, it provides a practical basis for predicting noise response from baseline performance.

\begin{figure}[htbp]
\centering
\includegraphics[width=\linewidth]{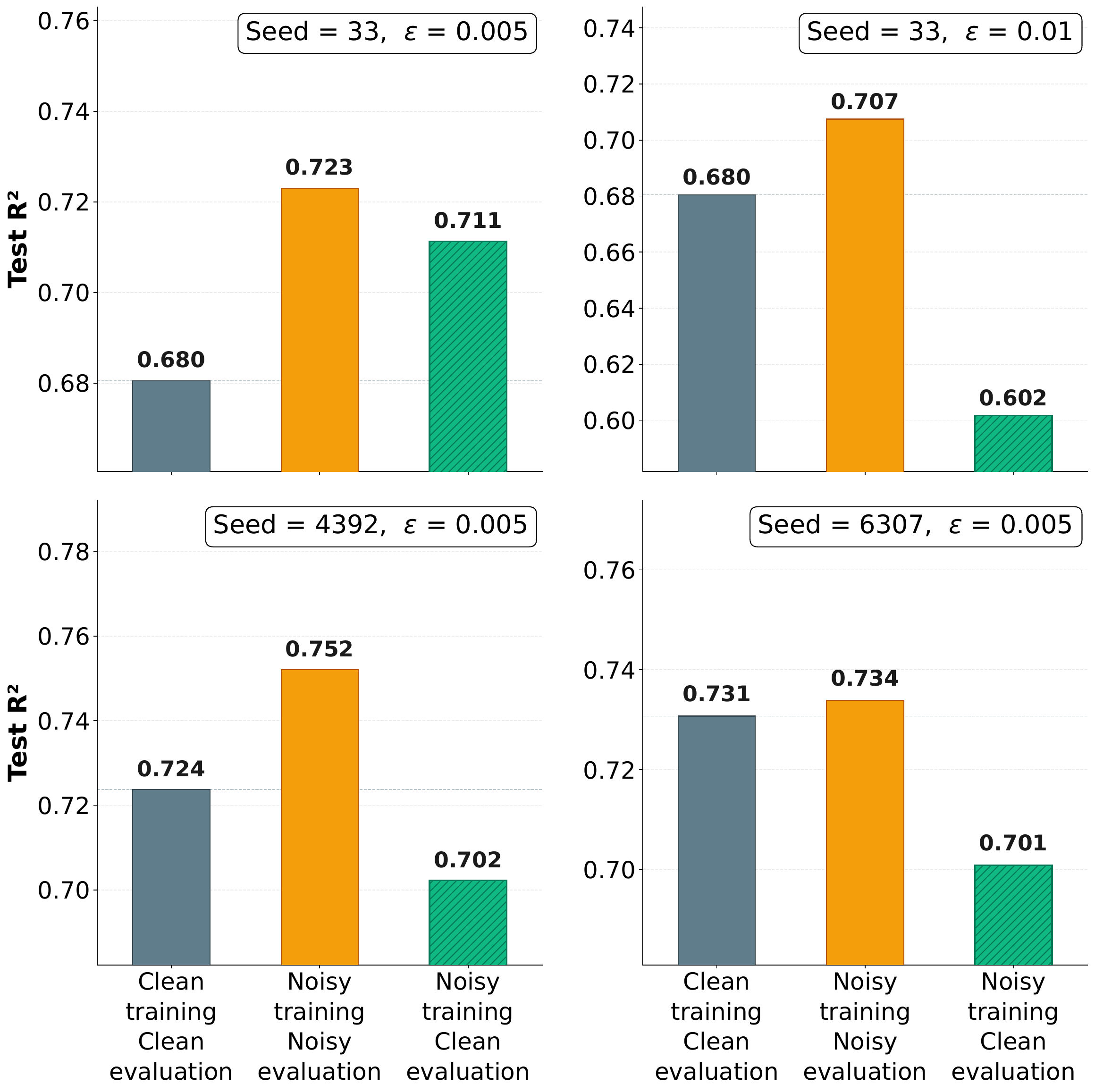}
\caption{\REV{Ablation study for three beneficial models tested at four noise conditions. Three settings are compared: clean training with clean evaluation (grey), noisy training with noisy evaluation (orange), and noisy training with clean evaluation (green). The dashed line marks the clean-trained baseline. Seed 33 at $\varepsilon = 0.005$ shows genuine optimization-time benefit, with the noisy-trained parameters retaining their advantage under noiseless evaluation. The remaining cases show improvement under matched noisy-training and noisy-evaluation conditions.}
} 
\label{fig:ablation_noise}
\end{figure}

\REV{ To probe the mechanism underlying the observed noise benefit, we performed an ablation study. Parameters obtained under noisy training $\theta_{\text{noisy}}$ were re-evaluated under noise-free conditions and compared with the clean-trained baseline and shown in \cref{fig:ablation_noise}.

The ablation reveals that the noise-benefit mechanism is itself initialization-dependent. For example, seed 33 at $\varepsilon = 0.005$ shows the noisy-trained parameters retained their advantage under noiseless evaluation. For seeds 4392 and 6307 at $\varepsilon = 0.005$, and for Seed 33 at $\varepsilon = 0.010$, the advantage was reduced under noiseless evaluation, indicating that the improvement is tied to matched noise conditions at inference.

On NISQ hardware, noise during inference is unavoidable. The matched noisy-training and noisy-evaluation condition therefore represents the actual deployment scenario, and adaptation to this condition is a meaningful form of noise-aware training. At the same time, the existence of genuine optimization-time benefit in some cases shows that this mechanism is not absent, but initialization-dependent.}

These findings \REV{are consistent with the interpretation} that quantum noise acts as an implicit regularizer, analogous to dropout or weight noise in classical neural networks. For under-optimized or weakly performing models, moderate noise appears to \REV{introduce stochastic perturbations into the optimization trajectory}, leading to systematic performance gains. In contrast, for well-converged models that have already reached near-optimal parameterizations, the same noise disrupts finely tuned representations and degrades predictive accuracy. This interpretation is consistent with classical regularization theory, in which the effectiveness of regularization depends on the degree of overfitting in the unregularized model. \REV{The observation that noise can enhance performance under specific regimes therefore suggests that, within the adopted effective noise model, noise need not be uniformly detrimental in quantum machine learning.}

\REV{ We also note that the present results do not by themselves establish that the observed improvement is unique to physical quantum noise. Since the adopted model introduces effective stochastic perturbations at the quantum-feature level, similar regularization-like behavior may also arise from more general stochastic optimization mechanisms. Therefore, we interpret the observed improvement as an effective-noise-induced regularization-like behavior, rather than as a quantum-noise-specific advantage.}

The observed regularization\REV{-like effect can be qualitatively interpreted} in terms of the geometry of the loss landscape. Quantum noise introduced during training \REV{may act} as a stochastic perturbation that alters the optimization trajectory. For poorly initialized or under-optimized models, this noise \REV{may help the model avoid unfavorable optimization trajectories and reach parameter regions with improved generalization}. In contrast, well-optimized models already reside in well-conditioned minima, where additional stochasticity primarily destabilizes finely tuned parameter configurations and therefore degrades performance. \REV{The ablation study in \cref{fig:ablation_noise} further shows that the optimization-time picture coexists with a distinct mechanism, in which noise benefits the model through adaptation to matched noise conditions rather than through improvements to the optimized parameters. }

\section{Conclusion and Perspectives}

\REV{By performing a systematic study of quantum graph neural networks for predicting molecular HOMO-LUMO gaps, we show that, within the adopted effective noise model, noise produces strongly heterogeneous effects across model instances.} Depending on the quality of the initial parameterization, noise can lead to either substantial performance gains or severe degradation. Across 55 random initializations, we find a clear and quantitative correlation between a model’s baseline performance and its response to noise: \REV{models with low noiseless baseline are more likely to improve under noise consistent with a regularization-like effect, whereas well optimized models with high noiseless baseline are more likely to lose accuracy.}

These heterogeneous responses add to growing evidence that quantum noise, when properly characterized and controlled, can play an active computational role rather than acting solely as a source of error. As quantum machine learning transitions from proof-of-concept demonstrations to practical workloads, algorithm design will need to account for the noise profile of the target hardware, adapting training and optimization strategies to exploit regimes in which noise improves generalization rather than degrades it.

Practical implementation of QGNNs on near-term quantum devices requires careful consideration of noise management strategies. The regularization effect identified in this work suggests a nuanced approach: rather than applying uniform error mitigation across all models, practitioners may benefit from tailoring noise management to model characteristics. For under-optimized models characterized by high initial loss, moderate noise levels may serve as an implicit regularizer that improves generalization without additional quantum resources.

Current benchmarking practices for variational quantum algorithms typically evaluate performance at fixed noise levels. However, the initialization-dependent response documented here suggests that such evaluations may be incomplete. A comprehensive assessment requires considering the distribution of initialization qualities and their corresponding noise responses. This perspective could reconcile conflicting reports in the literature regarding noise tolerance of quantum machine learning algorithms, where different studies may have inadvertently sampled from different regions of the initialization landscape.

This study examines single-layer QGNNs on molecular property prediction as a foundational case for understanding quantum noise effects. While this simplified setting reveals clear patterns, practical quantum algorithms often require deeper architectures where noise dynamics become more complex. Future investigations should explore how error accumulation and inter-layer correlations affect the regularization mechanism, potentially revealing optimal depth-noise trade-offs for different learning tasks.

Our phenomenological noise model provides valuable first-order insights but necessarily simplifies quantum error channels. The next critical step involves developing physically-motivated noise models that directly modify quantum channels rather than output distributions. Such models could incorporate device-specific error characteristics, including gate-dependent error rates, crosstalk, and temporal correlations, \REV{and would enable comparison with classical stochastic optimization baselines to clarify which aspects of the observed effect are specific to quantum noise.} By working at the quantum channel level, we could better understand how noise interacts with entanglement generation and quantum interference, fundamental resources in quantum computation that our current model does not explicitly address.

The heterogeneous response pattern discovered here, while statistically robust across 55 initializations, raises fundamental theoretical questions. Why does the transition between beneficial and detrimental responses occur at specific baseline performance values? What role does the quantum Fisher information, which quantifies the sensitivity of quantum states to parameter variations, play in determining noise susceptibility? Addressing these questions requires developing a comprehensive theoretical framework connecting initialization geometry, circuit expressivity, and noise resilience. Such theoretical advances could enable predictive models for noise response based on readily computable circuit properties.

Beyond QGNNs, the initialization-dependent noise effects likely extend to the broader landscape of variational quantum algorithms. Systematic studies across VQE for quantum chemistry, QAOA for optimization, and quantum kernel methods could reveal whether heterogeneous noise response represents a universal phenomenon or exhibits algorithm-specific characteristics. Understanding these patterns could inform the development of noise-aware ans\"atze and initialization strategies tailored to specific problem classes.

Finally, this work opens the door to treating quantum noise as a tunable hyperparameter rather than a fixed environmental constraint. Adaptive protocols that dynamically adjust noise levels during training, similar to learning rate scheduling in classical optimization, could optimize the exploration-exploitation trade-off throughout the learning process. Combined with improved theoretical understanding and hardware-aware modeling, such approaches could substantially enhance the practical utility of near-term quantum devices for machine learning applications.

\begin{acknowledgments}
The study of quantum machine learning in molecular science is supported by the Office of Naval Research (MURI Award 5 N00014-23-1-2001).
X.L. acknowledges additional support from the Center for MAny-Body Methods, Spectroscopies, and Dynamics for Molecular POLaritonic Systems (MAPOL) under subcontract from FWP 79715, which is funded as part of the Computational Chemical Sciences (CCS) program by the U.S. Department of Energy, Office of Science, Office of Basic Energy Sciences, Division of Chemical Sciences, Geosciences and Biosciences at Pacific Northwest National Laboratory (PNNL).
\end{acknowledgments}

\section*{Data Availability Statement}
The codes to reproduce the results presented in the paper are available on the GitHub repository:  \url{https://github.com/Linghua-Zhu/Quantum-Noise-QGNN}. These codes are released under the MIT license. 

\section*{Author Contributions}
L.Z., Y.D., and X.L. conceived the project. L.Z. and Z.Z. developed the computer code. X.L. acquired computing resources. L.Z. carried out the calculations and analyzed the data. X. L. acquired research funding. X. L., L.Z., Y.D. wrote the manuscript with input from all authors.  All authors discussed the results and approved the final version of the manuscript.

\section*{Competing Interests}
The author declares that there are no competing financial interests, personal relationships, or professional affiliations that could be perceived as influencing the work described in this manuscript.

\nocite{*}
\bibliography{aipsamp.bib}

\end{document}